\documentstyle{article}
\def\cl{\centerline}
\def\al{\alpha}
\def\be{\beta}
\def\de{\delta}
\def\la{\lambda}
\def\ra{\rightarrow}
\def\ur{\nearrow\ }
\def\ul{\nwarrow\ }
\def\dr{\searrow\ }
\def\dl{\swarrow\ }

\newcommand{\PSbox}[3]{\mbox{\rule{0in}{#3}\includegraphics{#1}\hspace{#2}}}

\def\bege{\begin{equation}}
\def\ende{\end{equation}}

\begin{document}
\begin{large}

\title{ 3-Dimensional MULTI-LAYERED 6-VERTEX  STATISTICAL MODEL: EXACT SOLUTION}
\author{ 
V. Popkov,\  Center for Theoretical
Physics, Seoul National University, Seoul 151-742  Korea\\
(  on leave from Institute for Low Temperature Physics,\\
 47 Lenin Ave,  310164 Kharkov, Ukraine)\\
B. Nienhuis, Institute for Theoretical Physics (University of
Amsterdam)\\
Valckernierstraat 65, NL 1018 XE Amsterdam}
\maketitle
\vskip 0.10cm

\newpage

\begin{abstract}
Solvable via Bethe Ansatz (BA) anisotropic statistical model
on cubic lattice consisting of locally interacting 6-vertex planes, is
studied. Symmetries of BA lead to infinite hierarchy of possible phases,
which is further restricted by numerical simulations.
The model is solved for arbitrary value of the interlayer coupling constant.
Resulting   
is the phase diagram in general 3-parameter space. Two new phases of 
chiral (spiral) character and new first order phase transition 
appear due to the interplane interaction.
Exact mapping onto the models with some inhomogenious 
sets of interlayer coupling
constants is established.
\end{abstract}
\vskip 0.20cm

\bigskip


\bigskip  

\section {Introduction}

Exactly solvable models e.g. models
for which set of physical quantities such as the bulk free energy,
the interfacial
tension, some critical exponents etc. can be calculated analytically,
play important role in statistical mechanics and the phase transitions
theory. Most solvable statistical models are two-dimensional
(see \cite{baxter} for review). YBE, or star-triangle equation,
serves as integrability
condition, or transfer-matrices commutativity condition
for $2D$ solvable models.
Unlike $2D$ models, solvable models in $3D$ are very rare examples.
Usually solvable $3D$ models --- see e.g.
 Zamolodchikov model \cite{zamolodch}
and its generalizations \cite{3Dmodels} --- are based on
solution of tetrahedron equation. The latter is natural generalization
of Yang-Baxter equation YBE and serves as  transfer-matrices
in $3D$ commutativity condition.

Unfortunately all  $3D$ solvable models known so far possess
common rather unsatisfactory
feature, from the viewpoint of applications to statistical mechanics
--- they incurably have negative \cite{zamolodch} or even complex
\cite{3Dmodels} Boltzmann weights.
Since these weights are probabilities (up to a normalization),this
property makes statistical mechanical interpretation of the models
highly problematical (although the associated quantum problem may still be
sensible).

Recently the method was proposed
in \cite{our,my} which allows to construct
solvable statistical models (however anisotropic ) with
positive Boltzmann weights and local interactions
in $3D$ starting from solvable $2D$
models. The idea is the following:
the whole row of infinite number of separate sites (vertexes) is
considered as a simplest object. The states probabilities (Boltzmann
weights) are defined as product of those for each site separately,
multiplied by interaction factors depending upon the local
configurations of each two neighbouring sites. Such construction
leads to multilayered models on $3D$-lattice, consisting of $2D$ integrable
planes-layers, with specific interaction between them. 
What is important
is that transfer-matrix commutativity condition turns out to be
not tetrahedron equation but an infinite set of usual Yang-Baxter
equations. These equations can be satisfied simultaneously,
producing the variety of $3D$-extended
models \cite{my} satisfying the necessary physical requirements --- positivity
of Boltzmann weights and locality of interactions.

Here we investigate the simplest example of such a model --- $3D$-extended
6-vertex model, with the homogenious set of interplane coupling 
constants.   
6-vertex model is rather popular object to study, because of
being limiting case for many solvable models in critical region
(see e.g. \cite{baxter}).
$3D$ solvable extension for 6-vertex model  was obtained in \cite{our}
and in \cite{my}
generalized onto other solvable vertex models with charge conservation
and inhomogenious sequences of interplane interaction constants.
In \cite{our} the phase diagram for the free fermions case was 
conjectured. 
 Here we go beyond this restriction and obtain the phase diagram in the
general 3-parameter space, for the extended 6-vertex model.

The paper is organized as follows: first we remind the definition
of the anisotropic $3D$-extended 6-vertex model and consider the strong 
interplane interaction limit. Then we give the alternative derivation
of Bethe Ansatz (BA) equations, using the established gauge equivalence
of our multilayer model to the set of $2D$ 6-vertex planes, each one 
in a field defined by the polarization of planes -- neighbours. Using
the symmetries of BA, we prove the equivalence between the model with 
homogenious set of interaction constants and some models with 
inhomogenious sets. We make use of these results, together with hypothesis 
of non-degeneracy of maximal eigenvalue, to eliminate the problem. 
The resulting phase diagram is obtained in the next section. Conclusion
and discussing the possible generalizations closes the paper.

\section { Definition of  the model}

The model we consider is system of $K$ planes. Each of these planes is the
symmetric 6-vertex solvable model on square $N\times M$ lattice \cite{liebwu}.
We can ''paste'' together $ij$-sites of all planes and formally get
$2D$ system with complex site consisting of $K$ simple vertexes
(see fig.~1).

\oddsidemargin -1cm
\topmargin -1cm
\textwidth 17cm
\textheight 23.2cm
\def\al{\alpha}
\def\be{\beta}
\def\de{\delta}
\def\la{\lambda}
\def\bege{\begin{equation}}
\def\ende{\end{equation}}

\begin{large}

\bigskip
\bigskip

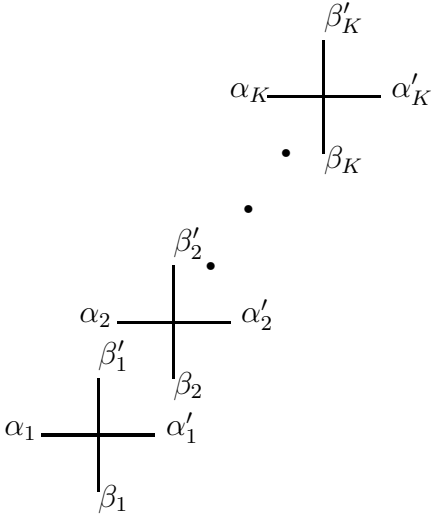
\begin{figure}

\setlength{\unitlength}{0.5mm}
\begin{picture}(90,120)(-15,-15)            
\thicklines
\put(-15,0){\line(1,0){30}}
\put(5,30){\line(1,0){30}}
\put(45,90){\line(1,0){30}}
\put(-25,0){$\al_1$}
\put(-5,30){$\al_2$}
\put(35,90){$\al_K$}
\put(18,0){$\al'_1$}
\put(38,30){$\al'_2$}
\put(78,90){$\al'_K$}
\put(0,-15){\line(0,1){30}}
\put(20,15){\line(0,1){30}}
\put(60,75){\line(0,1){30}}
\put(0,-19){$\be_1$}
\put(20,11){$\be_2$}
\put(60,71){$\be_K$}
\put(0,19){$\be'_1$}
\put(20,49){$\be'_2$}
\put(60,109){$\be'_K$}
\put(40,60){\circle*{2}}
\put(30,45){\circle*{2}}
\put(50,75){\circle*{2}}
\end{picture}

\caption{Multivertex consisting of $K$ simple vertexes }

\end{figure}

\end{large}

Boltzmann weight
of such multivertex fragment has form
\bege
{\bf L}_{\al'_1 \al'_2  \ldots \al'_K \be'_1
\be'_2 \ldots \be'_K}^
{\al_1 \al_2  \ldots \al_K \be_1 \be_2 \ldots \be_K}
 = \prod_{k=1}^K
 {L_{6v}}_{\al'_k, \be'_k}^{\al_k, \be_k}
{\exp\{ -h_k \al_{k} \be_{k+1} + h_k \al'_{k+1} \be'_{k} \}}
\label{bw}
\ende

where $h_k$ are arbitrary constants, defining interaction
between nearest neighbours in $k$-th and $(k+1)$-th plane,
$ {L_{6v}}_{\al'_k, \be'_k}^{\al_k, \be_k}$--- Boltzmann weights of ''source''
6-vertex solvable model, state variables $\al_k, \be_k ,\ldots$ take
values $\pm 1$.
For the planes (layers) we impose periodic boundary conditions $K+1 \equiv 1$.
The
permissible configurations of 6-vertex model are drawn on fig.2. The
variables sitting on edges are represented by arrows on fig.2. ''+''
(''-'') correspond to arrows pointing up or to the right
(down or to the left). The local Boltzmann weights are invariant
under inversion of all arrows.


\bigskip
\bigskip

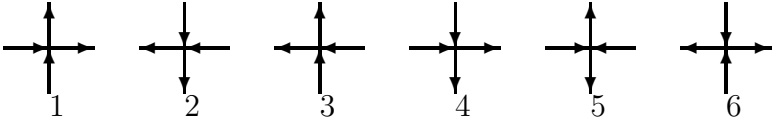
\begin{figure}
\setlength{\unitlength}{0.6truemm}
\begin{picture}(130,30)(-15,-15)            
\thicklines
\put(-15,10){\vector(1,0){10}}
\put(75,10){\vector(1,0){10}}
\put(105,10){\vector(1,0){10}}
\put(-5,10){\vector(1,0){10}}
\put(85,10){\vector(1,0){10}}
\put(145,10){\vector(1,0){10}}
\put(-5,0){\vector(0,1){10}}
\put(55,0){\vector(0,1){10}}
\put(145,0){\vector(0,1){10}}
\put(-5,10){\vector(0,1){10}}
\put(55,10){\vector(0,1){10}}
\put(115,10){\vector(0,1){10}}
\put(25,10){\vector(-1,0){10}}
\put(35,10){\vector(-1,0){10}}
\put(55,10){\vector(-1,0){10}}
\put(65,10){\vector(-1,0){10}}
\put(125,10){\vector(-1,0){10}}
\put(145,10){\vector(-1,0){10}}
\put(25,20){\vector(0,-1){10}}
\put(25,10){\vector(0,-1){10}}
\put(85,20){\vector(0,-1){10}}
\put(85,10){\vector(0,-1){10}}
\put(115,10){\vector(0,-1){10}}
\put(145,20){\vector(0,-1){10}}
\put(-5,-5){$1$}
\put(25,-5){$2$}
\put(55,-5){$3$}
\put(85,-5){$4$}
\put(115,-5){$5$}
\put(145,-5){$6$}
\end{picture}

\caption {Fig.2. Permissible vertex configurations for 6-vertex model. 
Boltzmann weights of configurations: $w_1=w_2=a;\ w_3=w_4=b;\ w_5=w_6=c$}
\end{figure}


\subsection* { Strong interplane interaction limit
$h \rightarrow \infty $.}

To understand the nature of interaction for the homogenious model
(all interplane interaction constants $h_k$ equal:
$h_k \equiv h$,  any $k$), let us consider the strong
interplane interaction limit $h \rightarrow \infty $.
The configuration with maximal Bolzmann weight will give the main
contribution to the partition sum. To obtain this configuration,
we should maximize exponential factor in (\ref{bw}) for all sites
in all planes.

Without loss of generality, we can set $\al_k = \be'_k = +1$ for $k$-th
plane. Then, to maximize the exponential factor in (\ref{bw}), we
should choose $\be_{k+1}= -1; \ \ \al'_{k+1} = +1$ in $k+1$-th plane,
i.e. vertex of type $4$ (see fig.2), and
$\al_{k+1} = +1;\ \  \be'_{k+1} = -1$. Performing the next step, we get
 $\be_{k+2} = \al'_{k+1} = -1$. This fits vertexes of type $5$ and
$2$; however vertex $5$ is unsuitable\footnote{
In order to get maximal Boltzmann weight, we have to construct
configuration in each plane with the same type of vertexes.
This can be done with type-$1-4$ vertexes but not with  type-$5$
vertexes -they always go in pairs with type $6$ vertexes. }
Proceeding analogously further, we get: planes $k - (k+4)$
are formed by the vertexes of type $1,4,2,3,1$, respectively.
It is easy to see that each next vertex (next plane) can be obtained
from the previous one by the clockwise rotation by $\pi/2$. After full
rotation over $2 \pi$, the configuration repeats.

Thus, in the strong interplane interaction limit, the model has
homogenious structure within each horizontal plane, and spiral
structure with period 4 in vertical direction: each next plane
configuration is obtained from the previous one by clockwise
$\pi/2$ rotation:

\bege
\ldots \ur   \dr    \dl   \ul   \ldots 
\label{ourNN}
\ende

The arrows in formula above are the resulting from vector summation
of all arrows in each plane: for example, the arrow $\ur$ corresponds to
homogenious configuration of vertexes of type 1 in Fig.2. 
 
\def\cl{{\cal L}}

\bigskip
\section { The matrix formulation}
\bigskip

It is convenient to rewrite (\ref{bw}) in matrix form.
Let us introduce $2^{2K} \times 2^{2K}$ matrix $\cl^k$ acting in the
tensor product $\prod_1^K{\otimes g_0} \prod_1^K{\otimes g_n}$
with elements
\begin{eqnarray}
& {(\cl^k)}_{\al'_1 \al'_2  \ldots \al'_K \be'_1
\be'_2 \ldots \be'_K}^
{\al_1 \al_2  \ldots \al_K \be_1 \be_2 \ldots \be_K} =
\de_{\al_1\al'_1} \de_{\be_1\be'_1}
\de_{\al_2\al'_2} \de_{\be_2\be'_2}
\ldots
\de_{\al_{k-1}\al'_{k-1}} \de_{\be_{k-1}\be'_{k-1}}
\exp {(- h_k \al_k \be_{k+1} + h_{k-1} \al'_k \be'_{k-1}  )}
\nonumber  \\
& {L_{6v}}_{\al'_k \be'_k}^{\al_k \be_k}
\de_{\al_{k+1}\al'_{k+1}} \de_{\be_{k+1}\be'_{k+1}}
\ldots
\de_{\al_K\al'_K} \de_{\be_K\be'_K}
\label{lbw}
\end{eqnarray}

${L_{6v}}$ is the $4\times4$ local $L$-matrix of six-vertex model
with the following nonzero elements:
\bege
L_{1 1}^{1 1}= L_{2 2}^{2 2} = a;
\ \  L_{1 2}^{1 2}= L_{2 1}^{2 1} = b;
\ \  L_{2 1}^{1 2}= L_{1 2}^{2 1} = c
\label{6vlm}
\ende
in which $a,b,c$
 are Boltzmann weights of permissible configurations (see fig.2).
Here and below in this section we borrow notations from paper of 
Faddeev and  Takhtajan
 \cite{Faddeev}.
Note that matrices ${\cl^k}$, corresponding to neighbouring $k$-s, do not
commute.
The multivertex Boltzmann weights (\ref{bw}) are matrix
elements of the ordered matrix product of $\cl^k$ over all planes:
$$\prod_{k=1}^K{\cl^k} = {\bf L} $$
\bege
{\bf L}_{\al'_1 \al'_2  \ldots \al'_K \be'_1
\be'_2 \ldots \be'_K}^
{\al_1 \al_2  \ldots \al_K \be_1 \be_2 \ldots \be_K}
 = \prod_{k=1}^K
{L_{6v}}_{\al'_k, \be'_k}^{\al_k, \be_k}
{\exp\{ -h_k \al_{k} \be_{k+1} + h_{k-1} \al'_{k} \be'_{k-1} \} }
\label{bw1}
\ende
(expressions (\ref{bw}) and (\ref{bw1}) are the same, due to
periodicity).

Using introduced notations we can write partition function of the model
in usual (see e.g. \cite{baxter}) form:
\bege {\bf Z} = Tr ({\bf T})^M
\nonumber
\ende
with transfer-matrix ${\bf T}$ being the trace of the ordered product of
local matrices ${\bf L}$ along the row (we'll denote matrix
${\bf L}$ acting in the site $n$ as ${\bf L}_n $):
\bege
{\bf T}= Tr{ (\prod_{n=1}^N{ \bf L}_n) }
\label{gtm}
\ende
Here $Tr$ operation  and matrix product goes over $\al$-indexes
 (we'll call it: 'auxiliary' space) and $Tr$ operation in the previous
formula (6) goes over $\be$-indexes ('quantum' space).

The free energy per site in the thermodynamic limit $N,M,K \ra \infty$
is defined by the maximal transfer-matrix eigenvalue
\bege
{\bf T} \Psi = {\bf \Lambda} \Psi; \ \ \ 
 f = - k_B T\ \lim_{N,K \rightarrow \infty} {1 \over NK}
  | {\bf \Lambda}_{max}|.\ \nonumber
\ende

Important feature of our $K$-plane model is that its monodromy matrix
${\cal T}$ can be
written as an ordered product of more simple ones  (we'll denote matrix
${ \cl^k}$ from (\ref{lbw}) acting in the site $n$ as $ \cl_n^k $):
\bege
{\cal T}
 =  \prod_{n=1}^N{ \bf L}_n =
 \prod_{n=1}^N  ( \prod_{k=1}^K{\cl_n^k} ) =
 \prod_{k=1}^K  ( \prod_{n=1}^N{\cl_n^k} ) =
\prod_{k=1}^K {\cal T}_k
\label{gmm}
\ende
\bege
{\cal T}_k =     \prod_{n=1}^N{\cl_n^k}
\label{lmm}
\ende
Matrix $\cl^k$ from (\ref{lbw}) can be written in compact form as can be
easily verified:
\def\sig #1{{\sigma^{(#1)}}}
\def\ta #1{{\tau^{(#1)}}}
\bege
\cl^k =
\exp{(-h_k \sig k \ta {k+1} )}
{L_{6v}}^k
\exp{(h_{k-1} \sig k \ta {k-1} )}
\label{mlbw}
\ende
where we denote by the $\sig k$ and $\ta k$  the diagonal matrix
$\sigma^z = diag(1,-1) $
acting nontrivially
in $k$-th  'auxiliary' and $k$-th 'quantum' space, respectively:

\bege
\def\sigmak{{
(\sig k)_{\al'_1 \al'_2  \ldots \al'_K \be'_1
\be'_2 \ldots \be'_K}^
{\al_1 \al_2  \ldots \al_K \be_1 \be_2 \ldots \be_K} =
\de_{\al_1\al'_1}
\ldots
\de_{\al_{k-1}\al'_{k-1}}
(\sigma^z)_{\al_k\al'_k}
\de_{\al_{k+1}\al'_{k+1}}
\ldots
\de_{\al_{K}\al'_{K}}
\prod_{i=1}^K
{\de_{\be_i\be'_i}}
}}
\def\tauk{{
(\ta k)_{\al'_1 \al'_2  \ldots \al'_K \be'_1
\be'_2 \ldots \be'_K}^
{\al_1 \al_2  \ldots \al_K \be_1 \be_2 \ldots \be_K} =
\de_{\be_1\be'_1}
\ldots
\de_{\be_{k-1}\be'_{k-1}}
(\sigma^z)_{\be_k\be'_k}
\de_{\be_{k+1}\be'_{k+1}}
\ldots
\de_{\be_{K}\be'_{K}}
\prod_{i=1}^K
{\de_{\al_i\al'_i}}
}}
\begin{array}{c}
\sigmak   \\
.         \\
\tauk
\end{array}
\label{sigmatau}
\ende

\bigskip

\bigskip

\section { Bethe Ansatz equations }

\bigskip

It can be shown - see Appendix for a proof - that the $3D$ model under
consideration (\ref{bw}) is gauge 
equivalent \footnote{We thank Yury Stroganov for drawing our attention
to this fact}
to the set of $2D$ 6-vertex planes,
each in its own horizontal field with the strength defined by
the vertical polarization in neighbouring planes and interplane interaction
constants:

for the $k$-th plane the field strength
\bege
H_k = h_k y_{k+1} - h_{k-1} y_{k-1}
\label{field}
\ende
where polarization $y_k$ is defined as usual:
\bege
y_k =
{n_k^\uparrow - n_k^\downarrow \over n_k^\uparrow + n_k^\downarrow } =
{2\; n_k^\uparrow - N \over N}
\label{yk}
\ende
$n_k^\uparrow\ \  ( n_k^\downarrow)$ is the number of upward (downward)
pointing arrows in horizontal row in plane $k$.

Therefore, the Bethe Ansatz of the model is given by 
the well-known formulas for 6-vertex
model in an external horizontal field (see e.g. \cite{liebwu}); below
$n_k = n_k^\uparrow$, $y_k = {2 n_k - N \over N} $):
\def\kb{{(k)}}
\def\tt{\tau_j^{\kb}}
\bege
 \Lambda_{k}(H_k) =  a^N  e^{N H_k} \prod_{j=1}^{n_k}
  {a \tt - b( 2 \Delta \tt - 1) \over a - b \tt}
  +
  b^N  e^{-N H_k} \prod_{j=1}^{n_k}
	{b - a( 2 \Delta  - \tt ) \over -a + b \tt}
\label{ev}
\ende
where $\tau_j^\kb$ satisfy the set of Bethe Ansatz equations
\bege
\def\ta#1{\tau_#1^{(k)} }  
e^{2 N H_k} ({\ta j})^N = (-1)^{n_k+1} \ \prod_{l=1}^{n_k}
{\ta j \ta l - 2 \Delta \ta j + 1 \over \ta j \ta l - 2\Delta \ta l + 1}
\label{ba}
\ende
$$  n_k = 1,2, \ldots N; \; \; \; k = 1,2 \ldots K$$
where
\bege  \Delta =  {a^2 + b^2 - c^2 \over 2 a b} \label{delta} \ende
 $a,b,c$ are Boltzmann weights of symmetrical
6-vertex model configurations (see fig.2).

The global transfer-matrix eigen-value is the product of those over all
planes:
\def\seth{\{ h \} }
\def\sety{\{ y \} }
\def\openseth{{h_1\ h_2 \ldots \ h_K}}
\def\opensety{{y_1\ y_2 \ldots \ y_K}}
\bege
 {\bf\Lambda}_{\sety}^{\seth} =
 {\bf\Lambda}_{\opensety}^{\openseth} =
 \Lambda_{y_1}(H_1)  \Lambda_{y_2}(H_2) \ldots
  \Lambda_{y_K}(H_K); \ \ \ y_p = {2 n_p - N \over N}
\label{gev}
\ende
 
The Eqs. (\ref{ev},\ref{ba})
were obtained directly
in \cite{our,my} using the quantum inverse scattering method,
and the analytic ansatz method, respectively.

We shall restrict ourselves to the model with all interplane 
interaction constants equal: $h_k \equiv h, \ \ any \ k$ (below we 
refer to that case as to the homogenious model). However as is shown
below, our results are valid also for some models with inhomogenious
sequences $\{ h_k \} $.

To obtain the bulk free energy and complete phase diagram, one should find
such a sequence of $n_k, \ \ \{ n_k \}_{k=1}^K$ that corresponds to 
the maximal transfer-matrix eigen-value $\Lambda_{max}$, and 
$\Lambda_{max}$ itself, in complete 3-parameter space $a/c,\ b/c,\ h$. 
This program
can be performed easily in the following limiting cases:

1) for the single 6-vertex model (decoupled planes limit $h = 0$). Then
BA solutions $\tau_i^{(k)}$ lie on a unit circle $\tau_j^{(k)} =
e^{i \la_j^{(k)}}$, $\la_j^{(k)}$ real. The expression for 
$\Lambda_{max}$ is found then by the integral equation method (see 
e.g. \cite{baxter}).

2)  'Ising chain' case  $\Delta = \pm \infty$, i.e. $a\  = \ 0$ or
$b\ =\ 0$, any $h$ is equivalent to  $\Delta = \pm \infty$, $h$ =0. Thus
it is reduced to the previous case.

3) for the free fermions case $\Delta = 0$. Then RHS of (\ref{ba}) is
equal to 1, and BA is solved trivially. The complete analysis is done
in \cite{our}. 

In the general case i.e. nonzero $h$ and $\Delta$, the structure of 
BA equations doesn't permit simple analysis. The reason is in that case
the locus of BA roots is unknown. The problem is similar to 
that arising when one considers usual 6-vertex model in external 
horisontal field
(see \cite{liebwu,baxter}).

\def\bl{{\Lambda}}
\def\boldl{{\bf \Lambda}}
\def\seth{{ \{ h \} }}
\def\sety{{\{ y \} }}
\def\openseth{{h_1\ h_2 \ldots \ h_K}}
\def\opensety{{y_1\ y_2 \ldots \ y_K}}

\bigskip
\centerline{\bf Connections between models with interplane constants}
\centerline{\bf  \{\ldots h,\ h,\ h, \ h, \ldots  \},  
		\{\ldots h,\ -h,\ h, \ -h, \ldots  \},  
		and arbitrary sets of 'h' and '{-h}'.}

\bigskip

The models listed in the title look different, and different they are,
having for instance, different 'strong interplane interaction limit'
$h \rightarrow \infty$ (this can be verified directly as it was done for
homogenious model at the beginning of the paper). We'll show however
that they have precisely the same transfer-matrix spectrum, and
therefore the same phase diagram, with minor redefinition of phases.

Let us take the state of the $k$-th plane with $n_k$ arrows pointing
up, in a field $H_k$, described by Eqs. (\ref{ev},\ref{ba}). Reversing
of all horizontal and vertical arrows, together with changing the sign
of $H_k$,  leaves the Boltzmann weights and therefore the eigen-value
(\ref{ev}) invariant:
\bege
\bl_{y_k} (H_k) = \bl_{-y_k} (-H_k) 
\label{reverse}
\ende
(Bethe Ansatz (\ref{ba}) changes accordingly).
Using this formula, write the global transfer-matrix eigenvalue
(\ref{gev}) for the set 
$\{ y_1, y_2, \ -y_3,\ -y_4, y_5, y_6, \ -y_7,\ -y_8,\ldots \ -y_K \} $
for the homogenious model:
\begin{eqnarray}
&\boldl_{ y_1\  y_2 \ -y_3 \ -y_4}^{h\ \ h\ \ h\ \ h} =
\bl_{y_1}(h(y_2 + y_K))
\bl_{y_2}(h(-y_3 - y_1))
\bl_{-y_3}(h(-y_4 - y_2))
\bl_{-y_4}(h(y_5 + y_3)) =  \nonumber \\
& \bl_{y_1}(h(y_2 + y_K)) \bl_{y_2}(h(-y_3 - y_1))
\bl_{y_3}(h(y_4 + y_2)) \bl_{y_4}(h(-y_5 - y_3)) \label{4period}
\end{eqnarray}

{\bf Remark.} 
Here and below we assume the number of planes $K$ to be infinitely
large and divisible by all numbers $K = 2*3*4*...$, $K+1 \equiv 1$,
to avoid complications connected with the boundary effects. 
For the sake of simplicity
we write down only the significant part of the multiplication (\ref{gev}),
then it continues periodically. 

Let us write down the global eigenvalue for the system with alternating 
constants:
\bege
 \boldl_{ y_1\  y_2 \ y_3 \ y_4}^{h\ \ -h\ \ h\ \ -h} =
 \bl_{y_1}(h(y_2 + y_K)) \bl_{y_2}(h(-y_3 - y_1))
\bl_{y_3}(h(y_4 + y_2)) \bl_{y_4}(h(-y_5 - y_3)) 
\nonumber
\ende
Comparing with the previous formula we have
\bege
 {\bf \bl}_{ y_1 \  y_2 \ -y_3 \ -y_4}^{h \ \ h \ \ h \ \ h} =
 {\bf \bl}_{ y_1 \  y_2 \ y_3 \ y_4}^{h \ \ -h \ \ h \ \ -h} 
\label{equiv1}
\ende
(periodical continuation is implied - see Remark above).

Analogously one obtains
\bege
 {\bf \bl}_{ y_1\  y_2 \ y_3 \ y_4}^{h\ \ h\ \ h\ \ h} =
 {\bf \bl}_{ y_1\  y_2 \ y_3 \ -y_4}^{h\ \ h\ \ -h\ \ -h} 
\label{equiv2}
\ende
Actually, one could coin such transformations between the model with
homogenious sequences and those with arbitrary sequences of $'h'$ and $'-h'$:
$ \{ h_k \}_{k=1}^K$, $h_k = \pm 1, \ \ \ k = 1,2, \ldots K$.
For instance, 
\bege
 {\bf \bl}_{ y_1\  y_2 \ y_3 }^{h\ \ h\ \ {-h}} =
 {\bf \bl}_{ y_1\  y_2 \ y_3 \ {-y_4} \ {-y_5}\ {-y_6}}^{h\ \ h\ \ h\ \ldots \ \ldots \ \ \ldots \  h} 
\label{equiv3}
\ende
and so on.

\section { Symmetries}

Here we make use of the results of previous paragraph to establish
transformations which map the eigenvalue to the eigenvalue of the
same model. 

First of all, note that simultaneous inversion of polarization
in all planes $Z$
\bege 
Z \sety = \{ -y \}
\label{Z}
\ende
leaves the eigenvalue invariant:
$$
\boldl_{\{ -y \} }^\seth =
 \prod_k \bl_{-y_k}^{h_k}(-h_k y_{k+1} + h_{k-1} y_{k-1}) = 
 \prod_k \bl_{y_k}^{h_k}(h_k y_{k+1} - h_{k-1} y_{k-1}) = 
\boldl_\sety^\seth 
$$
This is also the direct consequence of the fact that local Boltzmann 
weight (\ref{bw}) is invariant under transformation 
$ all \ \al_k,\be_k \rightarrow -\al_k, -\be_k $.

Another transformation we obtain, inverting the order of $\sety$-set
and $\{ -h \}$-set:
\def\invertsety{{y_1 \ y_K \ y_{K-1} \ldots \ y_3 \ y_2 }}
$$
\def\invertseth{{-h_K \ -h_{K-1} \ldots \ldots  -h_2 -h_1 }}
\def\isety{{ \ y_1 \ \ \ y_K \ y_{K-1} \ldots \ y_3 \ y_2 }}
\boldl_\opensety^\openseth = \boldl_\isety^\invertseth 
$$
(this is proved analogously)
Transformation J
\bege
J \{ \opensety \} = \{  \invertsety \}
\label{J}
\ende
is a symmetry for the model with alternating constants:
\def\setalth{{h \ -h \ h \ -h }}
$$ \boldl_\sety^\setalth = \boldl_{J \sety}^\setalth $$
Then, one more independent symmetry exist for that model:
\bege
F \sety  = \{ y_2 \ {-y_3} \ \ldots y_K \ {-y_1} \}; \ \ 
\boldl_\sety^\setalth = \boldl_{F \sety}^\setalth
\label{F}
\ende
For completeness, we define operator of one-step shifting $S$:
\bege
S \sety  = \{ y_{2} \ y_{3}\ldots \ y_K \ y_1  \}
\label{S}
\ende
Evidently, $S^n$, integer $n$, is a trivial symmetry for the homogenious model and $S^{2n}$
is a trivial symmetry for the model with alternating constants. 
Transformations $Z,J,F,S^2$ are basic symmetries for the model with
alternating constants. Let us obtain the basic symmetries for
the homogenious case. $Z$ and $S$ are the  symmetries already.

Then, define the transformation from (\ref{equiv1}) between
the two models:
$$ Q \sety = \{ y_1 \ y_2 \ {-y_3} \ {-y_4} \ldots \}; \ \ \ Q^{-1}= Q $$ 
Transformations $ Q J Q $ and $ Q F Q$ are the symmetries sought for:
\bege
P\sety = Q J Q \sety =
 \{ y_1 \ {-y_K} \ y_{K-1} \ {-y_{K-2}} \ldots y_3 \ {-y_2} \}; \ \ P^2 = I
\label{P}
\ende
$$
Q F Q  = S  
$$

So, for the homogenious model $all \ \ h_k \equiv h$ two nontrivial
symmetries $Z$ and $P$ exist defined by the (\ref{Z}),(\ref{P}). Working
out the same procedure for the model $h,\ h, \ -h, \ -h \ldots$ doesn't
give additional information. 
It is noteworthy also that the $Z$-symmetry is not independent but
can be expressed in terms of shifting and $P$-symmetry
$$ Z = (SP)^2$$
Nevertheless it proves convenient to keep it in mind as a separate symmetry.
As is shown below, $Z$ and $P$ symmetries,
having been combined with the non-degeneracy hypothesis, put drastic
constraints on the polarization set $\sety$ which corresponds to the
maximal eigenvalue of the global transfer-matrix of the homogenious model.

\def\bl{{\Lambda}}
\def\boldl{{\bf \Lambda}}
\def\seth{{ \{ h \} }}
\def\sety{{\{ y \} }}
\def\openseth{{h_1\ h_2 \ldots \ h_K}}
\def\opensety{{y_1\ y_2 \ldots \ y_K}}

\section { Finding the maximal eigenvalue set $\sety$ }

\def\y#1{{ \{y^{({#1})} \}    }}
\def\ytilde#1{{ \{\tilde{y}^{({#1})} \}    }}
We are now in a position to solve the maximal eigenvalue problem, for the
homogenious model. First we have to point out which set $\sety$
corresponds to it for any value of parameters $a/c, b/c, h$. We shall
parametrize the possible sets by a period $T$ and define the set
$\y T$ as that with periodically repeating entries:
$$ \y T = \{ y_1 \ y_2 \ldots \ y_T \ y_1 \ y_2   \ldots \ y_T 
\ldots \ y_T  \}; \ \ {\y T }_{T+n} ={\y T }_{n}
$$
Note that cases $T = 1$ and $T = 2$ are trivial because they lead to 
vanishing of interaction $h$ from all expressions (see (\ref{field})--
(\ref{ba})). The symmetry Z (see (\ref{Z})) does not effect the period $T$
of the set, but the symmetry (\ref{P}) does. Having been applied for
odd $T = 2 n +1 $, it doubles the period:
\bege
P \y T = \ytilde {2 T}
\label {doubling}
\ende
For instance for the period $T=3$ 
$$ P \{y_1 \ y_2 \ y_3 \ \ldots \} = 
 \{y_1 \ {-y_3} \ y_2 \ {-y_1} \ y_3 \ {-y_2} \ldots \}
$$
(We write down only the simplest periodically repeating pattern).

Note that set $\y T, \ \ odd \ T$ for the homogenious model,
corresponds to the sequence of period $4T$ for the model with alternating
constants: $Q \y T = \ytilde {4 T}$

For the sets with even $T = 2 n $ the symmetry (\ref{P}), generally 
speaking, does not effect the period, as can be easily verified.

To go further, we need an additional piece of information concerning the
maximal eigenvalue set $\sety$. We supply it by stating:

{\it The maximal eigen-value set for the homogenious model is unique, 
modulo shift, for all phases except the ferroelectric phase I 
(see fig.3). }

This statement is equal to stating that maximal global transfer
matrix eigenvalue is non-degenerate. For the 'source' 6-vertex model,
this is true: $y = 0$ for all phases except ferroelectric, and
$y=0$ is just the value invariant under action of $Z$-symmetry (\ref{Z}).
Strong coupling limit $ h \rightarrow \infty$ is non-degenerated, too.
As to the  exclusion - ferroelectric phase I (see fig.3), it has 
degeneracy $2^K$, $K$ being the number of planes. Indeed the ferroelectric
phase is built up from  $a$- (or $b$-) vertexes only. It follows from
(\ref{bw}) that Boltzmann weight does not depend on $h$ for the
homogenious model, whatever $a$-vertexes (type 1 or type 2 -see fig.2) 
form each plane. 

For all other phases, the hypothesis of non-degeneracy means that
the maximal set $\sety_{max}$ is invariant under action of $Z-$ and
$P-$symmetries (\ref{Z},\ref{P}), modulo arbitrary shift 
\footnote{
Note that for the model with alternating constants 
$\{\ldots h,\ -h,\ h, \ -h, \ldots  \}$, the maximal set 
$\sety_{max}$ is
always degenerate. It is the additional symmetry $F$ (\ref{F}) that 
accounts for the degeneracy. The same holds for the model with the
constants $\{\ldots h,\ h,\ -h, \ -h, \ldots  \}$.
}

The immediate consequences are:

1) from the (\ref{Z}): that the averave value of $< y_p >_{max}$ must
be zero:
$$ {1 \over K} \sum_{p=1}^K y_p = 0 $$

2) from the (\ref{doubling}): that the maximal set cannot have an odd 
period.

The period of maximal set also cannot be $T = 2 n, \  n \  odd$. To see
that, consider $T = 6$. For the set to be invariant under action of
$Z$-symmetry (\ref{Z}) it must have form
$$  \y 6 = \{ u \ v \ w \ {-u} \ {-v} \ {-w} \ldots \} $$
Acting on it by the $P$-symmetry 
$$ P \y 6 = \{ u \ w \ {-v} \ldots \}, $$
one obtains the set with period 3. Analogously one obtains 
$P \y {2n} = \ytilde n, \ \ n \ odd$, in contradiction with non-
degeneracy hypothesis. 

Since the maximal sets $\y T$ with $T$ and $T/2$ odd are forbidden,
we are left with the only possible choice
$$ T = 4 n$$
For that case, $P \y {4 n} = \ytilde {4 n}$ always.

The first nontrivial case is $T = 4$. Note that it is just the period
which arises in the strong interplane interaction limit $h \rightarrow 
\infty$. According to $Z$-symmetry (or as well to $P$-symmetry )
invariance, the maximal set reads
\bege
\y 4 = \{ \eta \ \xi \ {-\eta} \ {-\xi} \ldots \}
\label{T=4}
\ende
$T = 8$. $P$- and $Z$-symmetry invariance lead to the following set:
\bege
\y 8 = \{ u \ v \ 0. \ v \ {-u} \ {-v} \ 0. \ {-v}  \ldots \}
\label{T=8}
\ende
With increasing $T$, admissible structure looks more and more complicated.
For $T = 12 $
\bege
\y {12} = \{ u \ v  \ w \ r  \ {-w} \ v \ {-u} \ {-v} 
\ {-w} \ {-r} \ w \ {-v} \ldots \}
\label{T=12}
\ende
and so on. 
Thus, we have produced the hierarchy of sequences which are candidates
for the planes polarization set $\sety_{max}$ corresponding to the
maximal global transfer-matrix eigenvalue. The number of parameters is
reduced essentially but still there is quite a freedom left. The results
are in agreement with numerical data. The latter however forces us to
formulate the final hypothesis which we cannot prove. It is described
in the section to follow

\section { The final hierarchy }

Let us assume that the maximal set contains at least one plane $k$
with maximal polarization $y_k = 1$. It follows immediately
from (\ref{Z}) and non-degeneracy, that some another plane $p$ has
$y_p = -1$. Denote the distance between these two planes by $A = |p - k| $.
We shall parametrize the sequences $\sety$ by the value of $A$.
The entries of maximal set $\sety_{max}$ obey the following rule:

\bege
\left.
\begin{array}{cc}
 y_{k+n} = -y_{k-n}, \ \ &n \ odd   \\
 y_{k+m} = y_{k-m}, \ \ &m \ even   
\end{array}
\right]
\label{rule}
\ende
and the same with the replacement of $k$ by any other number $\tilde{k}$,
where $y_{\tilde{k}} = \pm 1$. From (\ref{rule}) those numbers are
$\tilde{k} = k \pm A n, \ integer \ n$. Making use of (\ref{rule}),
for $A$ odd we obtain the $y$-set of period $4 A$ $\y {4A}$, and
 $\y {2A}$, for $A$ even. 
The rules (\ref{rule}) are consistent with $Z$ and $P$
symmetries, and can be derived from them.

Let us take $A=1$. Here we have
\bege
\sety  = \{ 1 \ 1 \ {-1} \ {-1} \ldots \}
\label{A=1}
\ende
--- just the maximal set for the $h \rightarrow 
\infty$ limit (phase 4 in the phase diagram).
$A =2$ produces the set
\bege
\sety  = \{ 1 \ {-\xi} \ {-1} \ \xi \ldots \}
\label{A=2}
\ende
- the maximal set for the phase 5 in the phase diagram.
It coincides with that from (\ref{T=8}) when $ \eta = 1$.
For $A = 3$ one obtains from Eqs. (\ref{rule}): 
\bege
\sety = \{ 1 \ v  \ w \ {-1}  \ {-w} \ v \ {-1} \ {-v} \ {-w} \ 1 \ w \ {-v} \ldots \}
\label{A=3}
\ende
which coincides with (\ref{T=12}) when $u = 1, \ r = -1$.

$A = 4$ produces the set (\ref{T=8}) with the substitution $u = 1$, 
and so on.

Phases with $ A = 1,2$ do exist on the phase diagram. However we have
failed to find there numerically the phases with $ A = 3,4$. From that
we conclude that another members of hierarchy, with $A > 4$ don't 
appear, too. We suggest that the higher members of hierarchy
do appear when one includes into consideration more distant than
nearest neighbours, plane-plane interaction. For instance, we have
checked that the phase with $A=3$ (\ref{A=3}) does enter the phase
diagram when one includes additional interaction between the planes
$k \rightarrow k+3, \ \ any \ k$.

\def\bl{{\Lambda}}
\def\boldl{{\bf \Lambda}}
\def\seth{{ \{ h \} }}
\def\sety{{\{ y \} }}
\def\openseth{{h_1\ h_2 \ldots \ h_K}}
\def\opensety{{y_1\ y_2 \ldots \ y_K}}

\section{The phase diagram}

We have shown in the last two sections that the set of polarization 
constants $\sety = \opensety$ which corresponds to the maximal 
eigenvalue (the maximal set) must have the period divisible by 4:
$T = 4n$. For the simplest case $T=4$ the maximal set is (\ref{T=4})
and the transfer-matrix eigenvalue is given by Eq.(\ref{4period}),
with the substitution $y_1 = y_3 = y_5 = \eta; \ \ y_2 = y_4 = y_K = \xi$
$$ \boldl^{2/K} = \bl_\eta (2 h \xi) \bl_\xi (- 2 h \eta)$$
We can always choose $\xi$ and $\eta$ positive. Therefore we only
have to maximize the eigenvalue (\ref{gev}) within the two-parameter 
space 
$$\xi = { 2 n - N \over N}; \ \ \ \eta = { 2 m - N \over N}; \ \ \ 
n,m = 0,1 \ldots {N \over 2}
$$
By means of the Newton-Ralphson method we are able to solve BA equations
directly and find the block with largest eigen-value for system size
up to $N = 32$.

It turns out however that the maximal eigen-value for all values of
parameters $a,b,c$ and $h$ belongs either to the block 
$n = m = N/2$
(in that case $h$ vanishes from all equations (\ref{ev},\ref{ba}), 
yielding the known results
for 6-vertex model), or to the block with at least one of $n, m$, 
say $m$ equal to zero: $m = 0$. Finally 
the Bethe Ansatz (\ref{ev},\ref{ba}) rewritten
\def\tt{\tau_j}
\def\ds{\displaystyle}
\begin{eqnarray}
& \boldl^{2/K} = \bl_{m=0}(2 h \xi) \bl_\xi (2h) = 
\Lambda_1 \Lambda_2 \nonumber \\
& \def\kp{{2h(2n-N)}}
\Lambda_1 = a^N e^{\kp} + b^N e^{-\kp} \nonumber \\
& \ds{
 \Lambda_2 =
a^N e^{2hN} \prod_{j=1}^{n}
{a \tt - b( 2 \Delta \tt - 1) \over a - b \tt}
+ b^N e^{-2hN} \prod_{j=1}^{n}
{b - a( 2 \Delta  - \tt ) \over -a + b \tt}
}
\nonumber \\
& \ds{
 e^{4hN} ({\tau_j})^N = (-1)^{n+1} \ \prod_{l=1}^{n}
{\tau_j \tau_l - 2 \Delta \tau_j + 1 \over \tau_j \tau_l - 2\Delta \tau_l + 1}
}
\nonumber
\end{eqnarray}
Note that the eigen-value is invariant under the transformation: 
$$ a \rightarrow b, \  b \rightarrow a, \ h \rightarrow -h $$.

Figures \ref{Fig3a}-\ref{Fig3d} show 
the phase diagram of the model in the space of
Boltzmann weights ratios $a/c,\ b/c$, for the different values of
interplane constant $h$. Fig.3 (decoupled planes, $h$=0) repeats 
well-known results for usual 6-vertex model: the three distinct phases
exist separated by the lines $\Delta = \pm 1$.
These phases are (for details see \cite{liebwu}):

\begin{figure}
  \begin{center}
    \PSbox{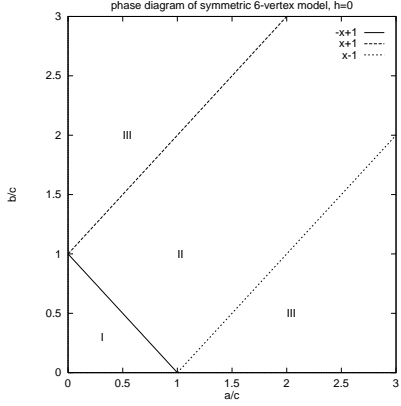   %
 voffset=0 hoffset=0 
angle=0 hscale=60 vscale=60}{11.5cm}{4.5cm}
  \end{center}
\caption{Phase diagram of the $2D$ 6-vertex model 
(corresponds to decoupled planes limit $h$ = 0)}
\label{Fig3a}
\end{figure}

I. Antiferroelectric phase. C-Vertexes (type 5 and 6) are dominant;
In sufficiently low temperatures vertex plane configuration is filled
with arrows alternating in both directions with step 1.

II. Disordered phase $-1 < \Delta < 1$. All types of vertexes are present;
there are no types of vertexes which are dominant. Here lies the high 
temperature limit so one should expect disorder.

III. Ferroelectric phase $\Delta > 1$. Phase III occupies two
separated regions, the left upper and the right bottom in the phase
diagram. In the left upper region, $b/c \ > \ a/c + 1$, each plane
is occupied by either exclusively type 3, or exclusively type 4
vertexes. It is convenient to consider the arrow resulting from
vector summation of all arrows in the plane:

 for the type 3 configuration, all arrows on the
plane are pointing up and to the left, resulting is upleft arrow $\ul$,
whereas for the type 4 configuration, the resulting is downright arrow
$\dr$, see Fig.2.  Polarization vector (\ref{yk}) modulus 
reaches its maximum $|y| = 1$, namely, $y = 1$ for the 
type 3 configuration and $y=-1$ for the type 4 configuration. 

If we associate with the each plane $k$ the corresponding arrow
--- $\ul$ for  type 3, $\dr$ for type 4 configuration,
then the state of the multiplane model in phase III,
left upper region, is characterized by the
random sequence
\bege
\ldots \ul \dr \dr \ul \dr \ul \ul \ul   \ldots
\label{III} 
\ende  
consisting of these arrows randomly placed. Of course switching on
some vanishing external field favouring type 3 configuration will
lift up this degeneracy, giving the homogenious sequence:
$$ \ldots \ul \ul \ul \ul \ul \ul \ul \ldots$$
The state of the model in the right bottom region, phase III,
$a/c \ > \ b/c + 1$ 
will be analogously characterized by the type 1, type 2 vertexes 
configurations, or upright, downleft $\ur , \ \  \dl$ arrows.
 $$ \ldots \ur \ur \ur \dl \ur \dl \dl \ur \ldots$$   
       
In Fig.\ref{Fig3b}- \ref{Fig3d} the phase diagrams for the systems with a fixed
$h$ in increasing order are given. The phases I - III are exactly
the same as in uncoupled planes  limit in Fig.\ref{Fig3a}.  
For nonzero coupling $h$ the regions occupied by the phase III are
given by the formula $ \Delta = \cosh{4h} $, and 
two new phases arise.

\begin{figure}
  \begin{center}
    \PSbox{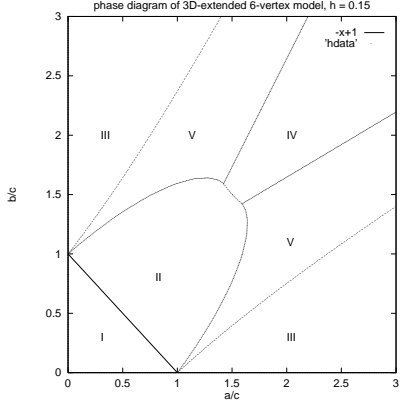   %
 voffset=0 hoffset=0 
angle=0 hscale=60 vscale=60}{11.5cm}{4.5cm}
  \end{center}
\caption{Phase diagrams of the 3D-extended 6-vertex model on cubic
lattice in the $a/c,\ b/c$ plane, $h$ = 0.15.  
The phase transitions between the phases II/IV, II/V, V/III are
 of the first order. The phase transitions between the phases IV/V are
 of the second order, Pokrovsky-Talapov type.
Transition I/II is the Kosterlitz - Thouless type transition. Curves
separating phases III/V , V/IV) are given by: $\Delta = \cosh {4h}$ 
$\ \  (b/c = (a/c) \  \exp{4h} -1$, respectively.}

\label{Fig3b}
\end{figure} 

\begin{figure}
  \begin{center}
    \PSbox{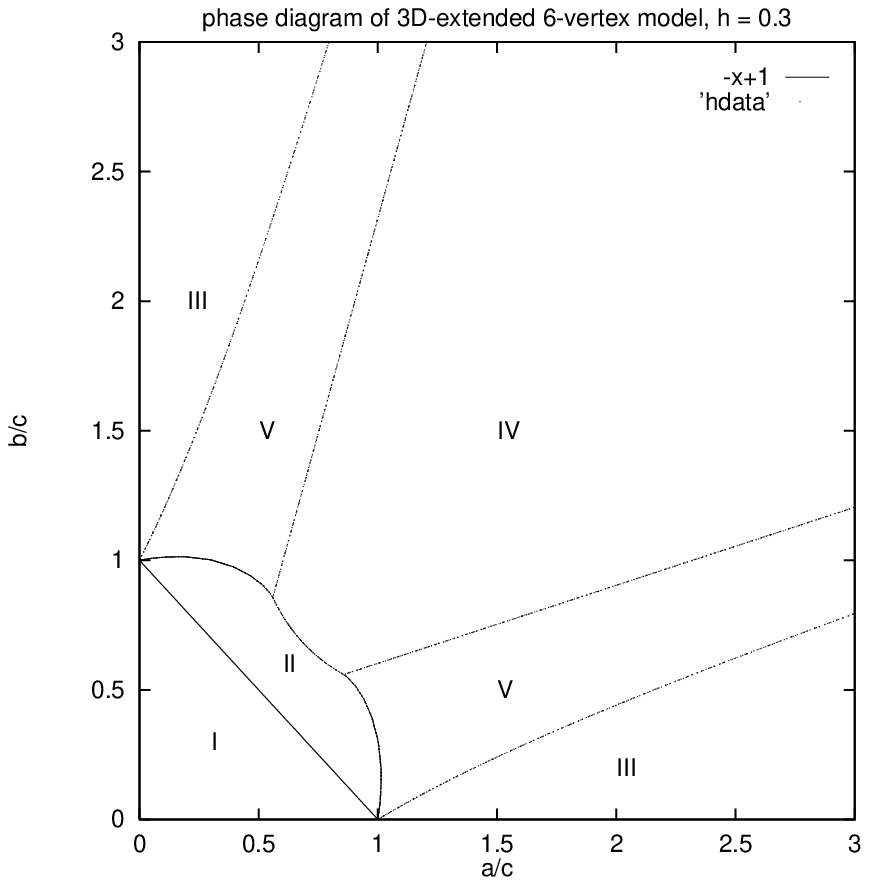   %
 voffset=0 hoffset=0 
angle=0 hscale=60 vscale=60}{11.5cm}{6.5cm}
  \end{center}
\caption{Phase diagram of the 3D-extended 6-vertex model on cubic
lattice in the $a/c,\ b/c$ plane, $h$ = 0.3}
\label{Fig3c}
\end{figure}

\begin{figure}
  \begin{center}
    \PSbox{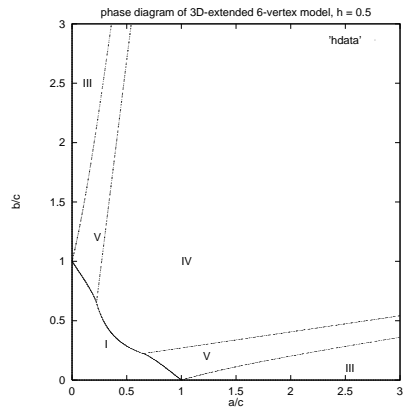   %
 voffset=0 hoffset=0 
angle=0 hscale=60 vscale=60}{11.5cm}{6.5cm}
  \end{center}
\caption{Phase diagram of the 3D-extended 6-vertex model on cubic
lattice in the $a/c,\ b/c$ plane, $h$ = 0.5}
\label{Fig3d}
\end{figure}

V. Layered antiferroelectric phase.  
This phase is exactly the one described  as the strong interplane
coupling limit in section 2. Each plane is in ferroelectric phase,
but plane $k+1$ configuration strictly follows from one in $k$-th 
plane by $\pi /2$ clockwise rotation. So, the planes-layers in
increasing order are formed by the unique, up to shifting, 
sequence of vertexes, or resulting arrows:
$$ \ldots  \ur \dr \dl \ul \ur \dr \dl \ul \ldots  $$
(compare with the sequence (\ref{III})). 
 
In that phase polarization 
vector (\ref{yk})
alternates each other plane  $y_{k +2} = - y_{k}$:
\bege
\ldots y_k, y_{k+1}, \ldots =
 \ldots  1;-1;-1;1;1;-1;-1;1 \   \ldots
 \label{0110}
\ende
forming the layered antiferroelectric structure (in each separate plane the
structure is ferroelectric)
Polarization vector modulus in each plane reaches its maximal value
$| y_k| = 1$, any $k$.

V. Phase V is an intermediate phase 
which can be described as follows:
 system splits into two
subsystems, each contains two planes $k$-th and $(k+2)$-th, e.g.
(the first and the third) and (the second and the fourth). In one of
the subsystems the planes are in ferroelectric phase, with alternating
polarization vector $ y_k = 1,\ \  y_{k+2} = -1$. Another subsystem
is partially disordered $y_{k+1} =  2 x - 1  = - y_{k+3} $.
Again as the previous one, this structure has period 4. 
Within the phase 4,
$x$ varies (see Fig. 4a,b); 
on transition line between phases IV/V $x$ goes to zero
continuously, forming the strong interplane coupling structure in phase 4. 
For the free fermions limit,
$\Delta = 0$, or $a^2 + b^2 = c^2$, value of $x$ is known
exactly (see \cite{our})
\bege
\cos \pi x = {a \over b} \sinh 4h,\ \  for\  a<b \nonumber
\ende

Outside the free fermions curve, the branches with equal polarizations
are arranged quite  regular, as is seen from Fig.\ref{Fig4a},\ref{Fig4b}. 
However we cannot present the exact formula for the moment.

\begin{figure}
  \begin{center}
    \PSbox{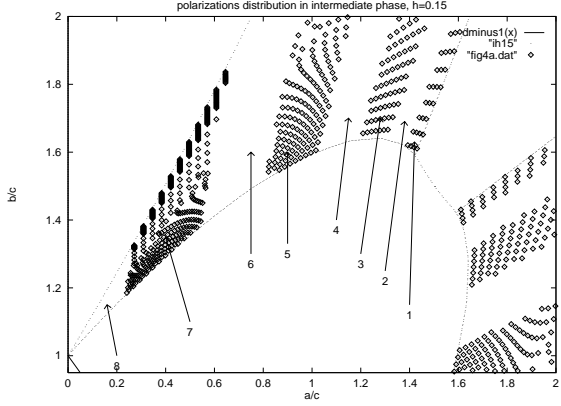   %
 voffset=0 hoffset=0 
angle=0 hscale=60 vscale=60}{11.5cm}{4.5cm}
  \end{center}
\caption{ Distribution of polarizations in the intermediate phase V. $h$=0.15.
This phase is characterized by the periodically repeated set of
polarizations (in planes) (\ref{A=2}) $\{ 1 \ {-\xi} \ {-1} \ \xi \ldots \} $,
$\xi = (2n-N)/N$, $n$ being the number of upward pointing arrows in a row
in the corresponding plane. The results of
numerical calculations of $\xi$ for the system size $N=32$ are shown. 
The area is divided into 8 sectors, sector number
$k$ corresponds to $2k-1 \leq n \leq 2k$. Thus, polarization 
monotonically increases from $\xi = -1$ on the second order phase
transitions line to $\xi =0$ near the point (0,1). On the 'free
fermions' circle $\Delta = 0$ inside the intermediate phase,
exact value of $\xi$ is given by $sin(\pi \xi/2) = -(a/b) sinh (4h)$.}
\label{Fig4a}
\end{figure} 

\begin{figure}
  \begin{center}
    \PSbox{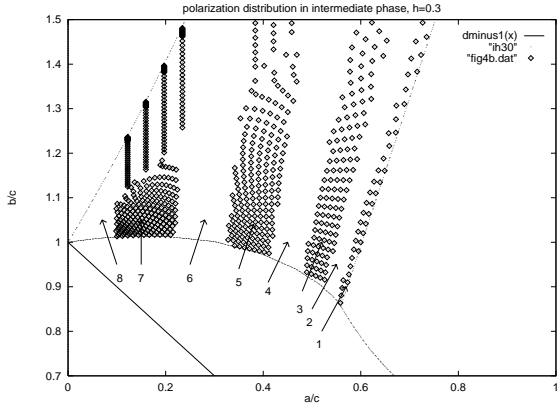   %
 voffset=0 hoffset=0 
angle=0 hscale=60 vscale=60}{11.5cm}{6.5cm}
  \end{center}
\caption{ Distribution of polarizations in the intermediate phase V. $h$=0.3.}

\label{Fig4b}
\end{figure}

The analytical and numerical calculations show that there are no other
phases at the phase diagram.  

The second order phase transition between the phases V/IV can be found 
considering the equality
\bege {\bf \Lambda} (n/N \rightarrow 0) = 1 \label{oneparticle} \ende.
Proceeding analogously as in \cite{liebwu} we get
$$ \tau_j = e^{-4h}$$.
For $a e^{4h} >b, b>a$ we have
$$ \prod {a \tt - b( 2 \Delta \tt - 1) \over a - b \tt} e^{-4h} =
 \prod {a  -  2 b \Delta  + b e^{4h} \over a e^{4h} - b }e^{-4h} = 1
$$
and ${a \over b} \sinh{4h} = 1 - \Delta e^{-4h}$, or using (\ref{delta}),
\bege
b/c = (a/c) \  \exp{(4h)} -1
\label{second1}
\ende
 All phase curves are symmetrical with respect to $a = b$ line.

Proceeding with the same equality (\ref{oneparticle}) and taking
$b>a e^{4h}$, we obtain: 
\bege
\Delta = \cosh{4h}
\label{ch4h}
\ende 
But now
this line marks only right margin for the phase transition III/V, and
not necessarily the exact point. We have found however with all numerical
accuracy that line (\ref{ch4h}) is exact transition point indeed.
With increasing $h$, the phases IV and V expand in space, and the others
diminish. When $h$ reaches $h_1 = 0.458134 \ldots$ and higher,
 phase II disappears completely 
 as is shown in Fig.\ref{Fig3d}. The fraction of space occupying by the phases
II,III, diminishes exponentially with $h$, as is seen from (\ref{second1},
\ref{ch4h}).

Finally, the transition line between the phases  II/IV (II/V) or I/IV (I/V)
in Fig.\ref{Fig3b}-\ref{Fig3d} is obtained 
numerically. The line ends in points (0,1) and
(1,0) which agrees with the limiting  'Ising chain' case. Phase transition
 II/IV (II/V) or I/IV (I/V) is of first order. Indeed in phases II or I
the partition sum does not depend on $h$ at all; (all $n_k \equiv N/2$ and
polarization vector $y_k$ for all $k$ is zero.  The values of 
 $y_k$-set jump when crossing the critical curve and dependence
on $h$ shows up. So the partition sum will have a cusp as a function
of $h$.

The transition IV/V is of the second order. On this line,
the order parameters --- polarization vectors $y_k$ (\ref{yk})
change continuously when approaching the critical point. The
second derivative for free energy over $h$ diverges as inverse
square root $\sim 1 / {\sqrt{h - h^*}}$ in the critical
point (Pokrovsky-Talapov type transition \cite{pokrovsky} ).

We have described the phase diagram of the $3D$-extended model with
the homogenious set of constants. The phase diagrams for the models
with arbitrary sets of $'h'$ and $'-h'$ constants 	
are the
same, with redefinition of phases (see Eqs.(\ref{equiv1}-\ref{equiv3})). 
Note that for the model with alternating constants,
$	\{\ldots h,\ -h,\ h, \ -h, \ldots  \}  $,
the period is always 2 (in planes).


\bigskip
\section* { Conclusion}
\bigskip

We have obtained the phase diagram for 3D solvable
multilayered 6-vertex model, in full 3-parameter space.
 The model enjoys locality of interactions
and positivity of Boltzmann weights. The applicability
of the method  to other solvable vertex models
with ice rule (\ref{ccc}) is shown in \cite{my}. 
in view of possible applications, note
that the strength of layer-
layer interaction $h$ can vary from plane to plane, as well as
anysotropy parameter within each layer. 
Another possibility is to include more distant than nearest neighbour,
interactions along 3-rd axis. The resulting solvable models
are ones with competing interactions \cite{nnn}.

Another interesting question is the universality class of the model we have
considered. The finite size scaling analysis (see e.g.\cite{Kim}) of new 
(due to plane-plane coupling) critical 
phase, named V in the phase diagram, shows that it is described 
by $2D$ conformal field theory with central charge $c = 1$. Thus it
belongs to the same  universality class as the ''source'' 6-vertex model.

\section*{ Acknowledgements}

One of the authors (VP) thanks colleagues from the Institute for
Theoretical Physics, University of Amsterdam, where this work was 
mainly done, for the hospitality, and Prof. Doochul Kim 
for nice introduction to finite size scaling.
  This work was supported
 in part by the INTAS grants 93 -- 1324, 93 -- 0633, 
Soros grant K5Z100 and by the Erwin  Shr\"odinger Institute, Vienna.
This work was supported in part by the Korea Science and Engineering
Foundation through the SRC program. We thank referees for good comments.

\def\bl{{\Lambda}}
\def\boldl{{\bf \Lambda}}
\def\seth{{ \{ h \} }}
\def\sety{{\{ y \} }}
\def\openseth{{h_1\ h_2 \ldots \ h_K}}
\def\opensety{{y_1\ y_2 \ldots \ y_K}}

\section*{Appendix }

The partition function for the system with open boundaries is given by
$$ {\bf Z} = {\cal T}^M $$
where  ${\cal T}$ is the global monodromy matrix. In our case 
(see Eqs. (\ref{lbw}-- \ref{mlbw})) ${\cal T}$
factorizes into product 
\bege
{\cal T} = \prod_{k=1}^K {\cal T}_k
\label{tau}
\ende
where the 'local monodromy matrix' for the $k$-th plane 
$$
{\cal T}_k = \prod_{n=1}^N 
e^{-h_k \sigma^{(k)} \tau_n^{(k+1)} } L_n^{(k)} 
e^{h_{k-1} \sigma^{(k)} \tau_n^{(k-1)}}
$$
$L_n$ being the matrix of Boltzmann weights for the 6-vertex model,
upper index $(k)$ corresponds to the $k$-th plane, matrices
$\sigma$ and $\tau$ are defined by (\ref{sigmatau}).
Omitting the index $k$ in the right-hand side of the last formula
for convenience and denoting 
$$ \sigma^{(k)} \rightarrow \sigma;\ \ \tau_n^{(k+1)}, \tau_n^{(k-1)}
\rightarrow \tau_n',\ \tau_n'', \ h_k \rightarrow h, 
\ h_{k-1} \rightarrow g,
$$
we have:
\def\ele #1{{ e^{-h \sigma \tau_#1'} \ L_#1 \  e^{ g \sigma \tau_#1'' } }}
\begin{eqnarray}
{\cal T}_k (h,g) = \prod_{n=1}^N \ele n =   \nonumber \\
\ele {1} \ \ele {2} \ldots \ele N
\label{tauk}
\end{eqnarray}
The exponential factors in this expression commute with each other
because they are diagonal matrices (see (\ref{sigmatau})). The commutation
with $L_n$ is given by 

$$
\lbrack \exp{( h {(\sigma+\tau_n)\tau_p'' )} )}, L_n \rbrack =
\lbrack \exp{( h {(\sigma+\tau_n)\tau_p' )} )}, L_n \rbrack = 0;
$$
which is equivalent \cite{our,my} to the charge conservation property
of the ''source'' 6-vertex model (\ref{6vlm}):
\bege {L_{6v}}_{\al'\be'}^{\al \be} = 0,\;\; unless\;\;
 \al'+\be' = \al + \be
\label{ccc}
\ende
Using these commutation rules, we move all exponents in (\ref{tauk}) 
outside to the left and to the right. For instance, to move the term
\def\est{{e^{-h \sigma \tau_2'}   }}
$\est $ to the left, one inputs unity
$$\est e^{\pm h \tau_1 \tau_2'} = \  e^{-h (\sigma +\tau_1) \tau_2'} \ 
 e^{h \tau_1 \tau_2'} $$
The first factor in this expression commutes with $L_1$ and goes to the
left while the second one commutes with all $L_1,\ L_2, \ldots L_N$
(because they act in the different subspaces) and goes to the right.
Repeating the similar procedure for all exponents, and taking 
into account ${\displaystyle \sum_{m=2}^N \sum_{n=1}^{m-1}  =
\sum_{n=1}^{N-1} \sum_{m=n+1}^{N} }$,
we obtain
\bege
\def\esum #1 #2{ e^{#1 \sigma \ \sum_{n=1}^N \tau_n{#2} } }
{\cal T}_k (h,g) = A_k \  
\esum {-h} {'} \ {\cal T}_k (0,0) \  \esum {g} {''} \ A_k^{-1}
\label{mm}
\ende
$$A_k = exp 
\left\{
\sum_{m=2}^N \sum_{n=1}^{m-1} ( -h \tau_m' \tau_n - g \tau_m \tau_n'')
\right\}
$$

Then, due to the charge conservation (\ref{ccc}) ,
${\displaystyle \sum_{n=1}^{N} \tau_n^{(k)}   }$ is a constant. For the 6-vertex
model this property is known also as the ice rule (see e.g. \cite{baxter,
liebwu}):
$$ {1 \over N} \ \sum_{n=1}^{N} \tau_n^{(k)} = y_k $$ --- the so-called 
polarization vector. $y_k$ may take values $-1 \leq \ y_k \ \leq 1$. 

Restoring the index $k$, $A_k$ can be written as
$$ A_k = C_{k+1,k} C_{k,k-1}; \  \ \ 
C_{k+1,k} = exp \left\{ 
\sum_{m=2}^N \sum_{n=1}^{m-1} (-h_k \tau_m^{(k+1)} \tau_n^{(k)} )
\right\}
$$
\def\tt #1{{ {\cal T}_{#1} }}
\def\ttt{{ {\cal T} }}
Multiplying $\tt 1 \tt 2 \ldots \tt K$ and using $A_k^{-1} \ A_{k+1} =
C_{k,k-1}^{-1} \ C_{k+2,k+1}$, we get for the global monodromy
matrix (\ref{gmm}):
\def\grob #1 #2 #3 {{
 \ e^{#3 \sigma^{(#1)} Y_{#2}} \ 
}}
\begin{eqnarray*}
\ttt = C_{1,0} C_{2,1} \grob 1 2 {-h_1} \tt 1 (0,0) \grob 1 0 {h_0} 
C_{1,0}^{-1} C_{3,2} \ldots \\
C_{K-1,K-2}^{-1} \ C_{K+1,K} \grob K {K+1} {-h_K}  \tt K (0,0)
\grob K {K-1} {h_{K-1}}
C_{K+1,K}^{-1} \ C_{K,K-1}^{-1}
\end{eqnarray*}
Here $ Y_k = N \ y_k$.

Moving $C$-factors ($C^{-1}$-factors) to the left (to the right), one obtains
\bege
\ttt  = B \left(
\prod_{k=1}^K  \grob k {k+1} {-h_k} \tt k (0,0) \grob k {k-1} {h_{k-1}}
\right) B^{-1}
\label{final}
\ende
$$B = \prod_{k=1}^K C_{k,k-1}$$

$B$ depends only on $\tau_n^{(k)}$, and therefore it is  a
gauge transformation. In the other hand, one can consider the transfer-matrix
for the 6-vertex model in a horizontal field of strength $H$
$$
\def\eh{e^{- {H \over 2 }  \sigma} }
\tt {6v} (H) = \prod_{n=1}^N \eh L_n \eh
$$
Doing the same procedure, we arrive to formula (\ref{mm}) with substitution
$h = -g = H/2; \ \ \tau_n',\tau_n'' \rightarrow 1$:
$$
\def\enh{e^{- {N H \over 2 }  \sigma} }
\tt {6v} (H) = A \ \enh \tt {6v} (0) \enh  \ A^{-1}
$$
in which $A$ again is a gauge transformation.
Comparing the last relation with (\ref{final}), we see that partition function 
of our model and the set 6-vertex planes, $k$-th plane in a field 
$$H_k = h_k y_{k+1} - h_{k-1} y_{k-1}$$
(both systems with open boundaries) are gauge-equivalent.  Hence,
their transfer-matrices are equal
$$
{\bf T}_{our \ model} = \prod_{k=1}^K T_{6v} (h_k y_{k+1} - h_{k-1} y_{k-1})
$$

\bigskip

\end{large}

\newpage



\bigskip
\begin{thebibliography}{99}
\bibitem{baxter} R.J. Baxter, Exactly solvable models in Statistical Mechanics
 (Academic Press, London, 1982)
\bibitem{lieb} E.H. Lieb, Phys. Rev.  {\bf 162}, 162 (1967)
\bibitem{zamolodch} A.B. Zamolodchikov , Soviet JETP {\bf 52}, 325 (1980);
A.B. Zamolodchikov, Comm. Math, Phys. {\bf 79}, 489 (1981)
\bibitem{3Dmodels} V.V. Bazhanov, R.J.Baxter, J.Stat.Phys. {\bf 69}, 453 (1992)
V. V. Mangazeev, S. M. Sergeev and Yu. G. Stroganov.
New series of 3D lattice integrable models.,
Int. J. Mod. Phys. {\bf A9}, 5517 (1994).
\bibitem{Faddeev}  L.A. Takhtajan and L.D.Faddeev, Uspekhi Mat. Nauk.
{\bf 34}:5 (1979) 13, translated in Russian Math. Surveys {\bf 34}:5 (1979) 
11.

\bibitem{our} A.E.Borovick, S.I.Kulinich, V.Yu.Popkov, Yu.M.Stzhemechny,
 Int. J Modern Phys. B, vol. 10, 443-453 (1996).
\bibitem{my} V.Yu. Popkov,   Anisotropic extension of 2D solvable models of statistics 
to 3D.//  Phys. Lett.A, {\bf 192}, 337 (1994).

 
\bibitem{nnn}  V. Popkov, V. Enolskii,  M. Salerno, 
{\it Exactly solvable multilayered 3D statistical model with competing 
interactions : connections with ANNNI model}, SNUTP 96-020 - preprint.
\bibitem{liebwu} E.H.Lieb, F.Y.Wu: in {\it"Phase  Transitions and Critical
Phenomena"},Vol.1 (Exact results). Ed. C.Domb, M.Green; Academic Press,
N.Y., 1972, p. 436.
\bibitem{pokrovsky} Pokrovsky V.L. and Talapov  A.L. 1980 Sov.
Phys. JETP {\bf 51}, 134.
\bibitem{Kim} J.D. Noh, D. Kim, Phys.Rev.E {\bf 53},3225 (1996).
\end{thebibliography}
\end{document}